\documentstyle[epsfig]{aipproc}

\begin{document}
\def\h{{\tiny \frac{1}{2}}}
\def\d{\delta}
\def\s{\sigma}
\def\p{\pi}
\title{Maximally-localized Wannier functions in perovskites:
Cubic BaTiO$_3$}

\author{Nicola Marzari and David Vanderbilt}
\address{Department of Physics and Astronomy, Rutgers University, Piscataway, 
NJ 08854-8019}

\maketitle

\begin{abstract}
The electronic ground state of a periodic crystalline solid is usually
described in terms of extended Bloch orbitals;
localized Wannier functions can alternatively be used.
These two representations are connected by families of 
unitary transformations, carrying a large degree of arbitrariness.
We have developed a localization algorithm that allows one to iteratively
transform the extended Bloch orbitals of a first-principles calculation into 
a unique set of {\it maximally localized} Wannier functions.
We apply this formalism here to the case of cubic BaTiO$_3$.
The purpose is twofold. First, a localized-orbital picture allows a 
meaningful band-by-band decomposition of the whole Bloch band complex. 
In perovskites, these Wannier functions are centered on the atomic 
sites and display clearly a $s$, $p$, $d$, or hybrid character.
Second, since the centers of the Wannier functions 
map the polarization field onto localized point charges, the ground state
dielectric properties become readily available. We study the Born 
effective charges of the paraelectric phase of BaTiO$_3$. We are able to 
identify not only the contributions that come from a given
group of bands, but also the individual contributions from the
``atomic'' Wannier functions that comprise each of these groups.
\end{abstract}

\section*{Introduction}

The electronic ground state of a periodic solid, in the independent
particle approximation, is naturally labelled according to the 
prescriptions of Bloch's theorem: single-particle orbitals are
assigned a quantum number {\bf k} for the crystal momentum, together
with a band index $n$. Although this choice is widely used in electronic 
structure calculations, alternative representations are available. 
The Wannier representation\cite{wannier}, essentially a real-space picture 
of localized orbitals, assigns as quantum numbers the lattice vector {\bf R}
of the cell where the orbital is localized, together with the band index
$n$. Wannier functions can be a powerful tool in the study of the electronic
and dielectric properties of materials: they are the solid-state equivalent
of ``localized molecular orbitals''\cite{boysr2}, and thus provide an 
insightful picture of the nature of chemical bonding, otherwise
missing from the Bloch picture of extended orbitals. 
In addition, the modern theory of polarization\cite{mtp} directly relates
the centers of the Wannier functions to the macroscopic polarization
of a crystalline insulator.

Wannier functions are strongly non-unique. This is a consequence
of the phase indeterminacy 
$e^{i\phi_n(\bf k)}$ that Bloch orbitals $\psi_{n{\bf k}}$ have at
every wavevector {\bf k}. This indeterminacy is actually more general than
just the phase factors: Bloch orbitals belonging to a 
composite group of bands (i.e. bands that are connected between themselves 
by degeneracies, but separated from others by energy gaps) 
can undergo arbitrary unitary transformations $U^{({\bf k})}$
between themselves at every {\bf k}. We have recently developed
a procedure\cite{marzari97} that can iteratively refine these otherwise
arbitrary degrees of freedom, so that they lead to Wannier functions
that are well defined and that  are localized around their centers 
(in particular, they minimize the second moment around the centers).
Such a procedure can be applied either to a whole band complex of Bloch
orbitals, or just to some isolated subgroups.

As a natural first application of this technique, we present here results
for the case of BaTiO$_3$ in the cubic phase. Perovskite ferroelectrics,
of which BaTiO$_3$ is a paradigmatic example, owe their very rich
phenomenology to the subtle competition of several degrees of freedom,
balancing the long-range dipole-dipole interaction with short-range Pauli 
repulsion. One of the striking features is the display of anomalously 
large Born effective charges\cite{zhong}. Their origin is
understood in a simple tight-binding picture\cite{harrison}: 
the change in the bond length (Ti-O in this case) corresponds 
to a dynamic charge transfer that is stronger when the bonding is 
borderline between ionic and covalent. Localized Wannier functions 
can thus be used to investigate the nature of this bonding, and to
monitor the changes that follow a ferroelectric distortion. Additionally,
the displacement of each Wannier center relates directly to the
effective charge contribution of its orbital, and can be used to identify
the nominal and the anomalous contributions to the polarization induced
by an atomic displacement.

\section*{Method}

Electronic structure calculations are carried out using periodic
boundary conditions. This is the most natural choice to study perfect
crystals and to minimize finite size-effects in the study of several
non-periodic systems (e.g. surfaces, or impurities). The one-particle
effective Hamiltonian $\hat{H}$ then commutes with the 
lattice-translation operator $\hat{T}_{\bf R}$,
allowing one to choose as common eigenstates the Bloch orbitals 
$\vert\,\psi_{n{\bf k}}\,\rangle$,
\begin{equation}
[\,\hat{H},\,\hat{T}_{\bf R}\,]\;=\;0 
\;\;\Rightarrow\;\; \psi_{n{\bf k}}({\bf r})\;
=\;e^{i\phi_n({\bf k})}\,u_{n{\bf k}}({\bf r})\,e^{i{\bf k\cdot r}}\;\;,
\end{equation}
where $u_{n{\bf k}}({\bf r})$ has the periodicity of the Hamiltonian.
There is an arbitrary phase $\phi_n({\bf k})$, periodic in reciprocal
space,
that is not assigned by the Schr\"odinger equation and that we
have written out explicitly.
We obtain a (non-unique) Wannier representation using any
unitary transformation of the form $ \langle\,n{\bf k}\,\vert\,{\bf
R}n\,\rangle=e^{i\varphi_n({\bf k})\,-i{\bf k\cdot {\bf R}}}:$
\begin{equation}
\vert\,{{\bf R}n}\,\rangle\;=\;\frac{V}{(2\pi)^3}\;\int_{BZ}\,
\vert\,\psi_{n{\bf k}}\,\rangle\,
e^{i\varphi_n({\bf k})\,-i{\bf k\cdot {\bf R}}}\,d{\bf k}\;\;. 
\end{equation}
Here $V$ is the real-space primitive cell volume.
It is easily shown that the $\vert\,{{\bf R}n}\,\rangle$ form
an orthonormal set, and that two Wannier functions 
$\vert\,{{\bf R}n}\,\rangle$ and $\vert\,{{\bf R}^\prime}n\,\rangle$
transform into each other with a translation of a lattice
vector ${\bf R}-{\bf R^\prime}$\cite{blount}.
The arbitrariness that is present in $\varphi_n({\bf k})$ [or 
$\phi_n({\bf k})$] propagates to the resulting
Wannier functions, making the Wannier representation non-unique.
Since the electronic energy functional in an insulator is also
invariant with respect to a unitary transformation of its $n$
occupied Bloch orbitals, there is additional freedom associated with
the choice of a full unitary matrix (and not just a diagonal one)
transforming the orbitals between themselves at every wavevector {\bf k}.
Thus, the most general operation that transforms the
Bloch orbitals into Wannier functions is given by
\begin{equation}
\vert\,{{\bf R}n}\,\rangle\;=\;\frac{V}{(2\pi)^3}
\;\int_{BZ}\,\sum_m\,U_{mn}^{({\bf k})}\,\vert\,\psi_{m{\bf k}}
\,\rangle\,e^{-i{\bf k\cdot {\bf R}}}\,d{\bf k}\;\;.
\label{eq:wannier}
\end{equation}
The Wannier functions $w_n({\bf r}-{\bf R})=\vert\,{{\bf R}n}\,\rangle$, 
for non-pathological choices
of phases, are ``localized'': for a ${\bf R}_i$ far away from
${\bf R}$, $w_n({\bf R}_i-{\bf R})$ is a combination of terms like
$\int_{BZ}\,\,u_{m{\bf k}}(0) e^{i{{\bf k}\cdot({\bf R}_i-{\bf R})}}\,
d{\bf k}$,
which are small due to the rapidly varying character of the exponential
factor\cite{blount}.

\subsection*{Maximally-localized Wannier functions}

Several heuristic approaches have been developed that construct 
reasonable sets of Wannier functions,
reducing the arbitrariness in the $U_{mn}^{({\bf k})}$ 
with symmetry considerations and analiticity requirements\cite{heuristic}, 
or explicitly
employing projection techniques on the occupied subspace spanned
by the Bloch orbitals\cite{projection}.
At variance with those approaches, we introduce 
a well-defined {\it localization criterion}, choosing the functional
\begin{equation}
\Omega=\sum_n\left[ \langle r^2\rangle_n- \bar{\bf r}_n^{\,2} \right]
\label{eq:omega}
\end{equation}
as the measure of the spread of the Wannier functions.
The sum runs over the $n$ functions $\vert\,{{\bf 0}n}\,\rangle$;
$\langle r^2\rangle_n$ and $\bar{\bf r}_n=\langle{\bf r}\rangle_n$
are the expectation values $\langle\,{{\bf 0}n}\,\vert\,
r^2\,\vert\,{{\bf 0}n}\,\rangle$ and
$\langle\,{{\bf 0}n}\,\vert\,{\bf r}\,\vert\,{{\bf 0}n}\,\rangle$.
Given a set of Bloch orbitals $\vert\,\psi_{m{\bf k}}\,\rangle$,
the goal is to find the choice of $U_{mn}^{({\bf k})}$ 
in (\ref{eq:wannier}) that 
minimizes the values of the localization functional 
(\ref{eq:omega}).
We are able to express the gradient $G=\frac{d\Omega}{dW}$ 
of the localization functional
with respect to an infinitesimal unitary rotation of our set of Bloch
orbitals
\begin{equation}
\vert u_{n\bf k}\rangle \;\rightarrow\; \vert u_{n\bf k}\rangle
+ \sum_m dW_{mn}^{\bf(k)} \, \vert u_{m\bf k}\rangle \;\;,
\label{eq:newu}
\end{equation}
where $dW$ an infinitesimal antiunitary matrix
$dW^\dagger=-dW$ such that
\begin{equation}
U_{mn}^{({\bf k})}=\delta_{mn}+dW_{mn}^{\bf(k)} \;\;.
\label{eq:rot}
\end{equation}
This provides
an equation of motion for the evolution of the $U_{mn}^{({\bf k})}$,
and of the $\vert\,{{\bf R}n}\,\rangle$ derived in (\ref{eq:wannier}),
towards the minimum of $\Omega$; small finite steps in the direction
opposite to the gradient decrease the value of $\Omega$, until a
minimum is reached.

\subsubsection{Real-space representation}

There are several interesting consequences stemming from the choice of
(\ref{eq:omega}) as the localization functional, that we briefly summarize
here.
Adding and subtracting the off-diagonal components 
$ \widetilde{\Omega}=\sum_n \sum_{{\bf R}m\ne{\bf
0}n} \Bigl\vert \langle{{\bf R}m}\vert{\bf r}\vert
{\bf 0}n\rangle\Bigr\vert^2 $, we obtain the decomposition
$\Omega= \Omega_{\rm\,I}+\Omega_{\rm D}+\Omega_{\rm OD},$
where $\Omega_{\rm\,I}$, $\Omega_{\rm D}$ and $\Omega_{\rm OD}$ are
respectively
$$
\Omega_{\rm\,I} = \sum_n \left[ \langle r^2\rangle_n- \sum_{{\bf R}m}
\,\Bigl\vert \langle{{\bf R}m}\vert{\bf r}\vert
{\bf 0}n\rangle\Bigr\vert^2 \right]\,,
$$
$$
\Omega_{\rm D} = \sum_n \sum_{\bf R\ne0}
\,\Bigl\vert \langle{{\bf R}n}\vert{\bf r}\vert
{\bf 0}n\rangle\Bigr\vert^2\,,
$$
$$
\Omega_{\rm OD} = \sum_{m\ne n} \sum_{\bf R}
\,\Bigl\vert \langle{{\bf R}m}\vert{\bf r}\vert
{\bf 0}n\rangle\Bigr\vert^2\,. 
$$
It can be shown that all terms 
are {\it positive-definite} (in particular $\Omega_{\rm\,I}$,
see Ref.\ \cite{marzari97});
more importantly,
$\Omega_{\rm\,I}$ is also {\it gauge-invariant}, i.e., it is invariant
under any arbitrary unitary transformation (\ref{eq:wannier})
of the Bloch orbitals. The
minimization procedure thus corresponds to the minimization
of $\widetilde\Omega = \Omega_{\rm\, D}+\Omega_{\rm OD}$. At the minimum,
the elements $\Bigl\vert \langle{{\bf R}m}\vert{\bf r}\vert
{\bf 0}n\rangle\Bigr\vert^2 $ are as small as possible, realizing the best
compromise in the simultaneous diagonalization,
within the space of the Bloch bands considered,
of the three position operators $x$, $y$ and $z$
(which do not in general commute when projected within
this space).

\subsubsection{Reciprocal-space representation}

As shown by Blount\cite{blount}, matrix elements of the position
operator between Wannier functions take the form
\begin{equation}
\langle{\bf R}n\vert{\bf r}\vert{\bf 0}m\rangle = i\,{V\over(2\pi)^3}
\int d{\bf k} \, e^{i{\bf k}\cdot{\bf R}}
\langle u_{n{\bf k}}\vert\nabla_{\bf k}\vert u_{m{\bf k}}\rangle 
\label{eq:rmatel}
\end{equation}
and
\begin{equation}
\langle{\bf R}n\vert r^2 \vert{\bf 0}m\rangle = -
{V\over(2\pi)^3} \int d{\bf k}\,
e^{i{\bf k}\cdot{\bf R}}
\langle u_{n{\bf k}}\vert\nabla_{\bf k}^2\vert u_{m{\bf k}}\rangle
\;\;.
\label{eq:rrmatel}
\end{equation}
These expressions provide the needed connection with our underlying Bloch
formalism, since they allow us to express the localization functional 
$\Omega$
in terms of the matrix elements of $\nabla_{\bf k}$ and $\nabla_{\bf k}^2$.
We thus determine the Bloch orbitals $\vert u_{m{\bf k}} \rangle$ 
on a regular mesh
of {\bf k}-points, and use finite differences to evaluate the above
derivatives. For any given {\bf k}-point in a regular cubic mesh 
(sc, fcc, bcc), we have a star {\bf b} of $Z$ {\bf k}-points that are first-neighbors;
their weights in the evaluation of derivatives are $w_b=3/Zb^2$.
We define $ M_{mn}^{(\bf k,b)} = 
\langle u_{m\bf k}\vert u_{n,\bf k+b}\rangle$ as
the matrix elements between Bloch orbitals at neighboring
{\bf k}-points. The $M_{mn}^{(\bf k,b)}$ are a central quantity in our
formalism, since we can then express all the contributions to the
localization functional using the connection 
made by Blount, together with our finite-difference
evaluations of the gradients. After some algebra we obtain\cite{marzari97}
\begin{equation}
\Omega_{\rm I} = {1\over N} \sum_{\bf k,b} w_b \,
\left( N_{\rm bands}-\sum_{mn} \vert M_{mn}^{(\bf k,b)} \vert^2 \right)\,,
\end{equation}
\begin{equation}
\Omega_{\rm OD}= {1\over N} \sum_{\bf k,b} w_b
\sum_{m\ne n} \vert M_{mn}^{\bf (k,b)} \vert^2\,,
\end{equation}
and
\begin{equation}
\Omega_{\rm D}= {1\over N} \sum_{\bf k,b} w_b
\sum_n \left( - {\rm Im}\,\ln M_{nn}^{\bf (k,b)}
 - \overline{{\rm Im}\,\ln M_{nn}^{\bf (k,b)}} \right)^2 \;\;.
\label{eq:omd2}
\end{equation}
From these, we can calculate the change in the localization functional
in response to an infinitesimal unitary transformation of the Bloch
orbitals, as a function of the $M_{mn}^{(\bf k,b)}$; 
once these steepest-descents are available, it is straightforward to
construct a procedure that updates the $U_{mn}^{({\bf k})}$ towards the
minimum of the functional.

\section*{Results: the case of BaTiO$_3$}

We study here the cubic phase of BaTiO$_3$, using a plane-wave total-energy
pseudopotential approach with the local-density approximation
to the exchange-correlation functional.
We use norm-conserving
pseudopotentials in the Kleinman-Bylander representation, 
with $q_c$ kinetic-energy tuning\cite{ming} for the oxygen atom
and a Troullier-Martins procedure\cite{tm}
for the titanium, to bring the cutoff convergence down to
900 eV. The $3s$ and $3p$ levels of titanium have been
included in the valence. The Brillouin zone is
sampled with a $4\times 4\times 4$ Monkhorst-Pack mesh; the 
lattice parameter used is 3.98\AA.

\subsection*{Wannier functions of cubic BaTiO$_3$}

\begin{figure} 
\begin{center} \leavevmode
\epsfxsize 5.5truein
\epsfbox{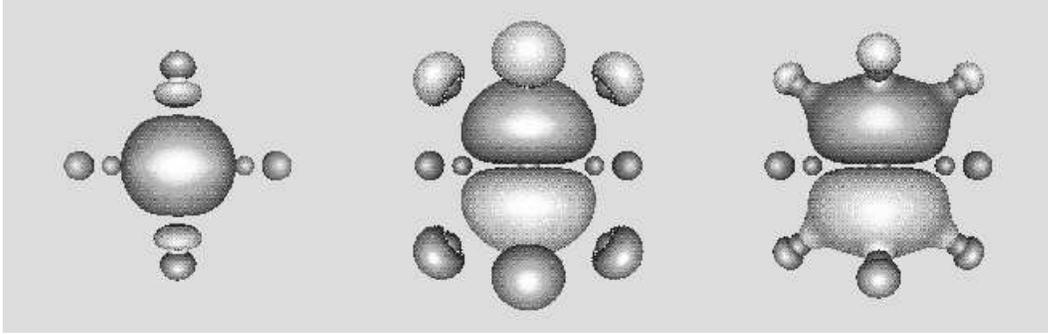}
\end{center}
\vspace{10pt}
\caption{Left panel: oxygen-centered Wannier function from the O
$2s$ 3-band group (the O atom is surrounded by four
Ba atoms on the sides, and two
Ti atoms on top and bottom).
Center and right panels: barium-centered Wannier function
from the Ba $5p$ 3-band group, and barium-centered Wannier function from the
Ba $5p$ and the O $2s$ 6-band group (the Ba atom is surrounded by 12
oxygens).}
\label{fig1}
\end{figure}

\begin{figure} 
\begin{center} \leavevmode
\epsfxsize 5.5truein
\epsfbox{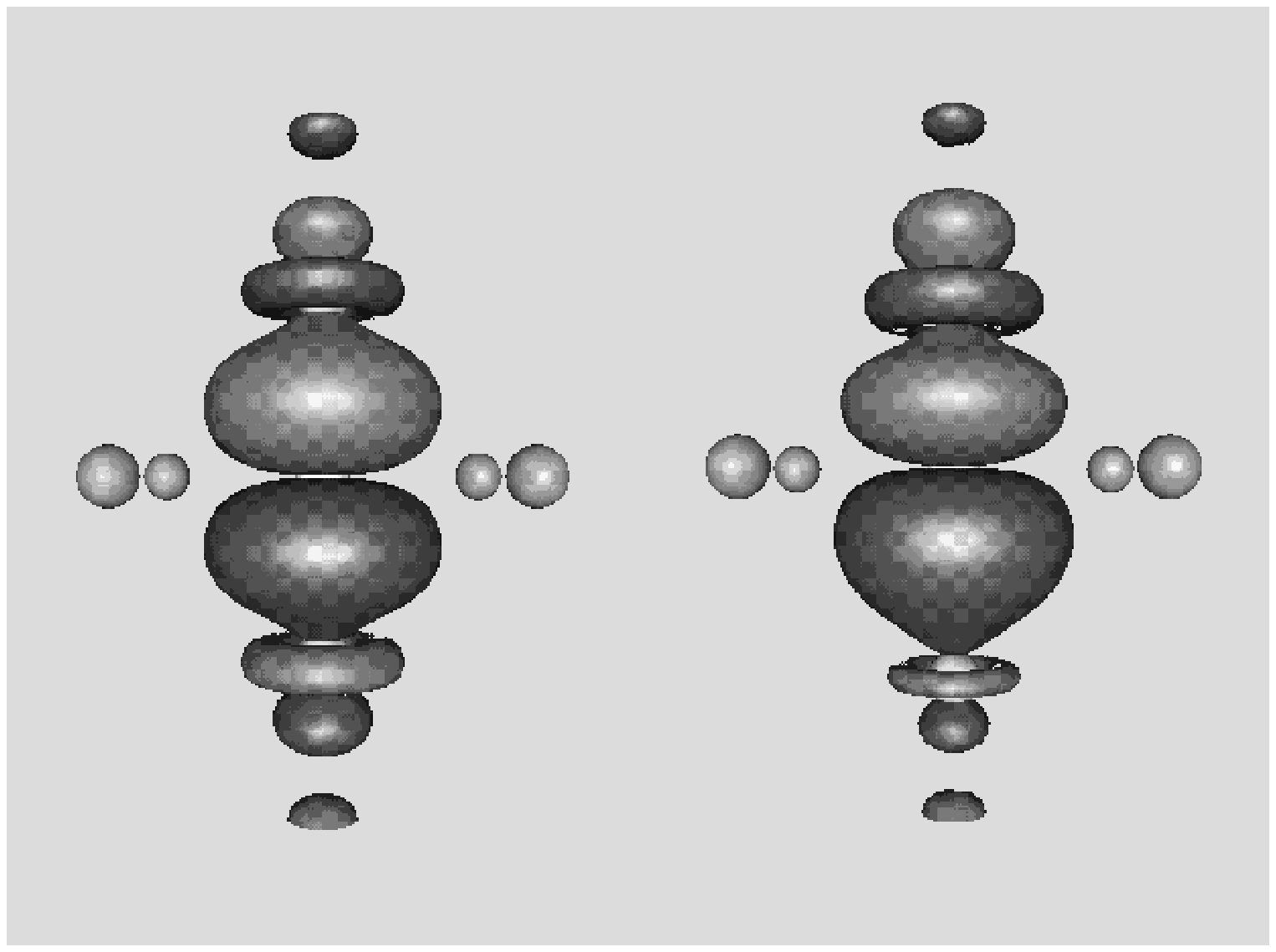}
\end{center}
\vspace{10pt}
\caption{Oxygen-centered $\sigma$ Wannier function from the localization of
the O $2p$ 9-band group. The orbital is oriented along the
Ti-O-Ti bond; the four Ba atoms neighboring the central oxygen and the 
two Ti atoms on top and bottom are also shown.
Left panel: ideal atomic positions. 
Right panel: same, but with the titanium atoms displaced downwards.}
\label{fig2}
\end{figure}

The minimization of the total energy provides the Kohn-Sham
Bloch orbitals on a regular mesh of {\bf k}-points, that are then used
as a starting point for the construction of the Wannier functions.
The subsequent minimization of the localization functional determines the
$U_{mn}^{({\bf k})}$ that correspond to the maximally-localized 
Wannier functions. In BaTiO$_3$ there are several groups of bands that are
separated by gaps. In order of increasing energy, we have the band
groups corresponding to the Ti $3s$ (1), Ti $3p$ (3),
Ba $5s$ (1), O $2s$ (3), Ba $5p$ (3) and O $2p$ (9) levels 
(in parenthesis are the number of bands in each group).
We initially consider each group of bands separately, and perform the
minimization on the 6 subspaces of dimensions
$1\times 1$, $3\times 3$, $1\times 1$,
$3\times 3$, $3\times 3$, and $9\times 9$.

The Wannier functions determined from the
the Ti $3s$ or the Ti $3p$ groups strongly resemble atomic orbitals,
slightly deformed by the underlying crystal potential,
and are not shown here.
We show instead in the left panel of Fig. \ref{fig1} one of the three
oxygen-centered Wannier functions that are derived from the oxygen $2s$
bands. In each unit cell there are three such functions, sitting on
each of the oxygens. The orbital shows its atomic $s$ character;
there are some contributions on the Ti,
with $p$ and/or $d$ character (the titanium is slightly
embedded in a $d_{z^2}$ orbital). Interesting results emerge
from the localization of the Ba $5p$ bands (Fig. \ref{fig1}, center and
right panels). These bands correspond to three orbitals in each unit cell,
all centered on the barium and oriented along the three cristallographic
directions ($p_x$, $p_y$ and $p_z$). 
It can easily be seen (center panel) that, in addition to
the distinctive atomic $p$ orbital on the barium, 
there are significant $sp$-like contributions
sitting on 8 of the 12 neighboring oxygens. This supports the suggestion
that barium in this compound has some covalent
character\cite{decomp}, and it is consistent 
with the anomalous effective-charge contributions
that come from this group of Wannier functions (see next subsection).
It is interesting to note that if we decide to treat the Ba $5p$ bands
together with the O $2s$ bands, we can (obviously) increase the degree
of localization of each orbital. In this latter case (right panel) the
$sp$ contributions on the oxygens decrease, being transferred to the 
$2s$ orbitals localized on the oxygens themselves.

Finally, we examine the 9 oxygen $2p$ bands that result from the
hybridization of the O $2p$ electrons with the Ba $6s$ and the 
Ti $4s$ and $3d$. There are three localized orbitals on each oxygen,
oriented along the Ti-O-Ti bonds. We label two of these orbitals
as $\pi$ and one as $\sigma$, according to their
symmetry along the Ti-O-Ti axis. One of the $\sigma$ orbitals is
shown in Fig.~\ref{fig2}, first with the atoms in their ideal positions
(left panel) and then with the Ti atoms displaced along the Ti-O bond (right
panel). The $\sigma$ orbital shows clearly the hybridization between
the oxygen $p$ orbital oriented along the Ti-O-Ti direction
and the $d_{z^2}$ orbital of the 
titanium. The $\sigma$ and even more the $\pi$ orbitals have
strong anomalous contributions to the effective charges, that can be
visualized with the large charge transfer from the oxygen 
atoms in response to the titanium displacements. 

\subsection*{Band-by-band decomposition of the Born effective charges}

\begin{table}
\caption{Born effective charges decomposed into atomic contributions, for
BaTiO$_3$.
The Z$^\star_{el}$ are calculated by displacing a Ba, Ti or a O
sublattice
(O$_1$ is parallel to the Ti-O direction, O$_2$ perpendicular) by $\delta
a_0$, where $\delta=0.002$.
The numbers in parenthesis are those of Ghosez et al. (a$_0$=3.94\AA).
Horizontal lines group band complexes that have been treated
together in the Wannier minimization. Top line in each group is
the total for that group.  We have added at the bottom
the O $2p$ decomposition obtained in a calculation performed
using a Ti atom with the $3s$ and $3p$ levels frozen in the core (Ti$^{4+}$).}
\begin{tabular}{ldddddddd}
                & Ba        & & Ti              & & O$_1$        & & O$_2$ & \\
                & $(\d,0,0)$ & & $(\h+\d,\h,\h)$ & & $(\h,\h,\d)$ & & $(\h+\d,\h,0)$ & \\ 
\tableline
Ti $3s$ (1)       &  0.01 &  (0.01) & -2.04 & (-2.03) &  0.03 &  (0.02) &  0.00 &  (0.00) \\ \tableline
Ti $3p$ (3)       &  0.02 &  (0.02) & -6.19 & (-6.22) &  0.21 &  (0.21) & -0.02 & (-0.02) \\
$(\h,\h,\h)$    &  0.00 & & -2.21 &          &  0.00 & &  0.00 & \\
$(\h,\h,\h)$    &  0.01 & & -1.99 &          &  0.21 & & -0.02 & \\
$(\h,\h,\h)$    &  0.01 & & -1.99 &          &  0.00 & &  0.00 & \\ 
\tableline
Ba $5s$ (1)       & -2.09 & (-2.11) &  0.04 &  (0.05) &  0.01 &  (0.01) &  0.02 &  (0.02) \\ \tableline
O $2s$ (3O)       &  0.65 &  (0.73) &  0.20 &  (0.23) & -2.45 & (-2.51) & -2.21 & (-2.23) \\
$(0,\h,\h)$     & -0.11 & &  0.48 &          & -0.04 & & -0.01 & \\
$(\h,0,\h)$     &  0.38 & & -0.14 &          & -0.04 & &  0.01 & \\
$(\h,\h,0)$     &  0.38 & & -0.14 &          & -2.36 & & -2.21 & \\ \tableline
Ba $5p$ (3Ba)     & -7.20 & (-7.38) &  0.31 &  (0.36) & -0.11 & (-0.13) &  0.50 &  (0.58) \\
$(0,0,0)$       & -2.46 & &  0.04 &          & -0.06 & &  0.21 & \\
$(0,0,0)$       & -2.37 & &  0.14 &          &  0.01 & &  0.01 & \\
$(0,0,0)$       & -2.37 & &  0.14 &          & -0.06 & &  0.29 & \\ \tableline
O $2p$ (9)        &  1.26 &  (1.50) &  3.01 &  (2.86) & -9.57 & (-9.31) & -6.35 & (-6.50) \\
$(0,\h,\h) \s$  & -0.01 & &  0.81 &          & -0.07 & &  0.03 & \\
$(0,\h,\h) \p$  & -0.20 & &  1.78 &          & -0.17 & & -0.49 & \\
$(0,\h,\h) \p$  & -0.20 & &  1.78 &          & -0.01 & &  0.00 & \\
$(\h,0,\h) \p$  &  0.22 & & -0.17 &          & -0.01 & &  0.01 & \\
$(\h,0,\h) \p$  &  0.51 & & -0.27 &          & -0.17 & &  0.04 & \\
$(\h,0,\h) \s$  &  0.10 & & -0.24 &          & -0.07 & &  0.02 & \\
$(\h,\h,0) \p$  &  0.22 & & -0.17 &          & -3.10 & & -1.89 & \\
$(\h,\h,0) \s$  &  0.10 & & -0.24 &          & -2.86 & & -1.81 & \\
$(\h,\h,0) \p$  &  0.51 & & -0.27 &          & -3.10 & & -2.26 & \\ \tableline
Total Z$^\star_{el}$ & -7.35 & & -4.67 &          & -11.87& & -8.06 & \\ 
Core            & 10.00 & & 12.00 & &  6.00 & &  6.00 & \\ \tableline
Total Z$^\star$ &  2.65 &  (2.77) &  7.33 &  (7.25) & -5.87 & (-5.71) & -2.06 & (-2.15) \\
Ref. \cite{zhong} &  2.75 & &  7.16 & & -5.69 & & -2.11 & \\ \tableline
Ti$^{4+}$ & & & & & & & & \\
O 2p (9)        &  1.29 &  (1.50) &  2.33 &  (2.86) & -8.93 & (-9.31) & -6.36 & (-6.50) \\
$(0,\h,\h) \s$  & -0.02 & &  0.49 & & -0.13 & &  0.01 & \\
$(0,\h,\h) \p$  & -0.19 & &  1.48 & & -0.19 & & -0.41 & \\
$(0,\h,\h) \p$  & -0.19 & &  1.48 & & -0.02 & & -0.01 & \\
$(\h,0,\h) \p$  &  0.22 & & -0.16 & & -0.02 & &  0.01 & \\
$(\h,0,\h) \p$  &  0.51 & & -0.25 & & -0.19 & &  0.04 & \\
$(\h,0,\h) \s$  &  0.11 & & -0.15 & & -0.13 & &  0.03 & \\
$(\h,\h,0) \p$  &  0.22 & & -0.16 & & -2.88 & & -1.89 & \\
$(\h,\h,0) \s$  &  0.11 & & -0.15 & & -2.50 & & -1.87 & \\
$(\h,\h,0) \p$  &  0.51 & & -0.25 & & -2.88 & & -2.28 & \\ 
\end{tabular}
\label{tab}
\end{table}

The Born dynamical effective charges describe the change in
macroscopic polarization that is induced by the 
displacement of a given ion. As such, they play a fundamental role in
determining the dynamical properties of insulating crystals, and are a
powerful tool to investigate the dielectric and ferroelectric properties
of materials. They also determine the splitting of the infrared-active
optical modes; in simpler compounds (e.g. GaAs) they can be
unambiguously determined from the experimental phonon dispersions.
Perovskite ferroelectrics display anomalous, large effective charges, that
can be almost double their nominal ionic value. The origin of this
effect lies in the large dynamical charge transfer that takes place when
moving away from the high-symmetry cubic phase (i.e., going from more ionic
to more covalent bonding). Orbital hybridization is necessary for
this transfer to take place; for this reason, our localized-orbitals
approach provides an insightful tool in examining these effects.
In the language of the modern thory of polarization\cite{mtp}, the anomalous
contribution is determined by the relative displacement of the Wannier
centers with respect to the ion that is being moved. If the bonding were
purely ionic, electrons (and thus Wannier centers) would be firmly
localized on each anion, and move rigidly with it. This is not the case 
in perovskite oxides. The anomalous contribution is often traced\cite{resta}
to substantial hybridization between the oxygen $p$ orbitals 
and the $d$ orbitals of the atom in the B site (Ti, in this case).
The picture can be somewhat more complex, with other group of bands playing
a role in the anomalous dielectric behavior.

We present in Table \ref{tab}
a full decomposition of the effective charges in BaTiO$_3$
coming from the different groups of bands; and, inside each group,
coming from the individual
Wannier functions identified by the localization procedure.
We compare the results for the groups of bands 
with those obtained by Ghosez et al. (Ref. \cite{decomp}).
We find very good agreement, given the difference in the
pseudopotentials used and
our choice of lattice
parameter. These results underline the conclusion
that the decomposition into band groups is
consistently defined in the linear-response formalism only
when the calculations are performed in the so-called 
{\it diagonal gauge}\cite{decomp}.
The effective-charge tensor reduces
to a scalar for barium and titanium, while it is diagonal with two 
inequivalent components (those parallel and perpendicular to the 
Ti-O bonds) for the oxygens.
Several facts stand out from an inspection
of the table. The anomalous effective charges originate not only
from the oxygen $2p$ bands; there are also sizeable contributions 
originating from the barium $5p$ orbitals and even from the oxygen $2s$.
More notably, there is a wide range of {\it compensating effects}
between groups of bands and between different orbitals inside each group. 
The partial cancellation of such large 
orbital polarizabilities hints again at the complexity of these materials,
which exhibit such a wide range of equilibrium properties as
a consequence of the existence of many of these competing effects.
Titanium shows the strongest deviations from a naive ionic picture.
The $\sigma$ and $\pi$ oxygen orbitals carry a {\it positive} electronic 
Z$^\star$ contribution, equal respectively to $0.81$ and $1.78$.
It should be noted that it is always the O $2p$ $\pi$ orbitals that 
carry the largest anomalous charge.
The O $2p$ contributions to the O$_1$ effective charges
(i.e.\ in the direction of the Ti-O bond) are also anomalous, 
up to $-1.10$ for each $\pi$ orbital (in addition to the nominal
$-2.00$ for each orbital).

Finally, we present at the bottom of Table \ref{tab}
the oxygen $2p$ decomposition
performed in a calculation where the $3s$ and $3p$ orbitals 
of the titanium have been removed from the pseudopotential (i.e.,
removed from the valence and frozen in the core).
It is interesting to note that most
contributions are completely unchanged; differences arise only for 
the anomalous contributions for the Ti and the O$_1$ displacements
(whose anomality, incidentally, is reduced if this more approximate 
formalism is employed).

\section*{Conclusions}

We have summarized here our formalism for obtaining
maximally-localized Wannier functions from the Bloch orbitals of
an ab-initio electronic structure calculation. This formalism can be very
helpful in understanding the chemical and dielectric properties of
materials. Perovskite ferroelectrics are a particularly promising class 
of systems to be studied, since the nature of the bonding and
hybridization can have a striking influence on the dielectric
properties and on the development of ferroelectricity. At variance
with other approaches, our method allows for a decomposition of
electronic properties (e.g., the effective charges) into meaningful atomic
contributions. In the case of BaTiO$_3$, it elucidates in particular
the origins of the large anomalous contributions to the effective charges.
\section*{Acknowledgments}

This work was supported by ONR Grant N00014-97-1-0048, and
NSF Grants DMR-96-13648 and ASC-96-25885.
We would like to thank Ph.~Ghosez and
X.~Gonze for providing an early copy of their work on the effective-charge
decompositions.

\end{document}